\documentclass[preprint,12pt,authoryear]{elsarticle}

\usepackage{amssymb}
\usepackage{graphicx}
\usepackage{braket}
\usepackage{amsmath}
\usepackage{hyperref}
\usepackage{float}
\usepackage{amsthm}
\usepackage{longtable}
\usepackage{booktabs}
\usepackage{CJKutf8}
\usepackage[strings]{underscore}
\usepackage{braket}
\usepackage{enumitem}
\usepackage{diagbox}

\makeatletter
\def\ps@pprintTitle{%
\let\@oddhead\@empty
\let\@evenhead\@empty
\let\@oddfoot\@empty
\let\@evenfoot\@oddfoot
}
\makeatother

\begin{document}
\begin{CJK}{UTF8}{gbsn}
\begin{frontmatter}



\title{Solving the Independent Domination Problem by Quantum Approximate Optimization Algorithm}

\author[a]{Haoqian Pan}
\author[a]{Changhong Lu}
\affiliation[a]{organization={School of Mathematical Sciences,  Key Laboratory of MEA(Ministry of Education) \& Shanghai Key Laboratory of PMMP, East China Normal University}, 
            city={Shanghai},
            postcode={200241},
            country={China}}

\begin{abstract}

In the wake of quantum computing advancements and quantum algorithmic progress, quantum algorithms are increasingly being employed to address a myriad of combinatorial optimization problems. Among these, the Independent Domination Problem (IDP), a derivative of the Domination Problem, has practical implications in various real-world scenarios. Despite this, existing classical algorithms for IDP are plagued by high computational complexity, and quantum algorithms have yet to tackle this challenge. This paper introduces a Quantum Approximate Optimization Algorithm (QAOA)-based approach to address the IDP. Utilizing IBM's qasm_simulator, we have demonstrated the efficacy of QAOA in solving IDP under specific parameter settings, with a computational complexity that surpasses that of classical methods. Our findings offer a novel avenue for the resolution of IDP.


\end{abstract}


\end{frontmatter}


\section{Introduction}\label{sec:Introduction}

The Independent Domination Problem (IDP) addresses the identification of the smallest independent dominating set (IDS) in a graph \(G(V,E)\). An IDS is a subset \(D\) of vertices such that every vertex in \(V \setminus D\) is adjacent to at least one vertex in \(D\), and no two vertices in \(D\) are adjacent. This problem is particularly relevant in applications such as wireless network design. Research has established the NP-completeness of the IDP across various graph classes, including but not limited to chordal, bipartite, circle, and triangle graphs \citep{RN447, RN449, RN445, RN444}. Given the complexity of the problem, the focus has shifted toward approximation and heuristic solutions, adapting to specific graph structures to enhance performance and feasibility. Significant contributions include the development of a polynomial-time algorithm with a time complexity of \(O(|V|^2)\) specifically tailored for permutation graphs \citep{RN442}. Furthermore, for cocomparability graphs, an advanced algorithm has been devised with a computational complexity of \(O(|V|^{2.376})\), representing a strategic optimization for this class of graphs \citep{RN443}. Additional strategies have been applied to at-most-cubic graphs, further illustrating the diversity of graph-specific solutions \citep{RN441}. For graphs without specific structural constraints, recent advancements propose algorithms with exponential time complexities such as \(2^{0.529|V|}\), \(2^{0.441|V|}\), and \(2^{0.424|V|}\), revealing a trend toward more efficient exponential algorithms \citep{RN451, RN452, RN450}. Despite these innovations, the challenge of reducing the algorithmic complexity of the IDP remains formidable, prompting ongoing research and development efforts in the field.

In recent years, the advancement of quantum computing technologies \citep{RN421,RN439,RN454} has significantly propelled the application of quantum algorithms to address combinatorial optimization problems. Among these algorithms, the Quantum Approximate Optimization Algorithm (QAOA) \citep{RN436} stands out as a prominent representative of the last decade, showcasing numerous applications at the intersection of combinatorial optimization and quantum computing. A variety of combinatorial optimization problems have been explored using QAOA on quantum computers and simulators. Notable examples include the Max-Cut problem \citep{RN436}, the Traveling Salesman Problem \citep{RN453}, the Domination Problem \citep{RN415}, the Minimum Vertex Cover Problem \citep{RN334}, the Boolean Satisfiability Problem \citep{RN455}, and the Graph Coloring Problem \citep{RN456}, among others. Despite the extensive body of research surrounding quantum computing and QAOA, there remains a notable gap in the literature regarding the application of these quantum algorithms to IDP. Specifically, the potential for utilizing QAOA or other quantum algorithms to effectively solve IDP remains unexplored. As such, the feasibility and efficacy of quantum-based solutions for this particular problem are yet to be determined.

The primary contribution of this paper lies in our pioneering application of the QAOA to solve the IDP, marking the first instance of utilizing a quantum algorithm for this specific challenge. We conducted comprehensive evaluations of the effectiveness of QAOA in addressing the IDP, focusing on two key aspects: fundamental testing and robustness testing. The findings presented herein offer valuable insights and foundational experience that will inform and guide future endeavors aimed at solving the IDP using quantum computing technologies.

The structure of this paper is organized as follows: In Section \ref{sec:problemmodeling}, we systematically introduce the methodology for transforming the IDP into a 0-1 integer programming model, followed by the formulation of a Quadratic Unconstrained Binary Optimization (QUBO) model and its corresponding Hamiltonian. Section \ref{sec:QAOA} presents an overview of the QAOA algorithm and delineates the procedural steps required to apply this algorithm for solving the IDP. Subsequently, in Section \ref{sec:Experiment}, we conduct both fundamental and robustness testing to assess the effectiveness of the QAOA-based algorithm in addressing the IDP. This section also includes a complexity analysis of the QAOA algorithm. Finally, Section \ref{sec:conclusion} provides a comprehensive summary of the paper.

\section{Problem modeling}\label{sec:problemmodeling}

When employing quantum algorithms to tackle combinatorial optimization problems, a prevalent strategy involves modeling or transforming the original problem into a QUBO format, which is subsequently converted into the Hamiltonian of the Ising model. Accordingly, in this chapter, we begin with a formal definition of the IDP and progressively derive the Hamiltonian representation pertinent to this problem.

\subsection{problem $\to$ 0-1 integer programming model}

Before delving into the IDP, we begin by providing a precise definition of the dominating set. Given a graph \( G = (V, E) \), a dominating set \( D \) is defined as a subset of \( V \) such that for every vertex \( v \in V \), the closed neighborhood \( N[v] \) intersects with \( D \) (i.e., \( N[v] \cap D \neq \emptyset \)). In this context, \( N[v] \) denotes the closed neighborhood of vertex \( v \). The objective of the domination problem is to identify the smallest such dominating set \( D \). The IDP represents a variant of the domination problem, which imposes an additional constraint: the dominating set \( D \) must also be independent, meaning that no edges exist between vertices in \( G[D] \). To facilitate the application of the QAOA in addressing this problem in the subsequent sections, we first present a mathematical model of the IDP.

\begin{alignat}{2}
  \min_{\{X_{i}\}} \quad & \sum\limits_{i=1}^{|V|} X_{i}  \\
  \mbox{s.t.}\quad
  &\sum\limits_{j \in N[i]} X_{j} \ge 1 \quad \forall i \in V  \label{const:i1} \\
  &X_{i} \in \{0,1\}  \quad \forall i \in V \\
  &X_{i} \cdot X_{j} = 0 \quad \forall ij \in E \label{const:i3}
\end{alignat}

In this model, \( X_{i} \) serves as a binary variable, where \( X_{i} = 1 \) indicates that vertex \( i \) is included in the dominating set \( D \), and \( X_{i} = 0 \) otherwise. Consequently, the expression \( \sum_{i=1}^{|V|} X_{i} \) represents the total number of vertices in the dominating set. Our objective is to minimize the size of \( |D| \), which is mathematically expressed as \( \sum_{i=1}^{|V|} X_{i} \). Constraint \ref{const:i1} ensures that for every vertex \( v \in V \), either \( v \) itself or one of its neighbors is included in \( D \). This constraint is a fundamental requirement of the domination problem. Constraint \ref{const:i3} introduces a unique condition for the IDP. The edges in graph \( G \) can be classified into three categories: (1) edges within \( G[D] \), (2) edges within \( G[V \setminus D] \), and (3) edges connecting a vertex in \( D \) to a vertex in \( V \setminus D \). For the second and third categories, since at least one vertex in each edge has \( X_{i} = 0 \), Constraint \ref{const:i3} is inherently satisfied. However, for edges within \( G[D] \), Constraint \ref{const:i3} necessitates that no edges exist within \( G[D] \), thus ensuring the independence condition is met. At this point, we have successfully established the 0-1 integer programming model for the IDP. The next step involves transforming this model into a QUBO model, followed by converting the QUBO model into a Hamiltonian representation.

\subsection{0-1 integer programming model $\to$ QUBO model}

The QUBO model is expressed in Eq. \ref{eq:quboform}, where \( x \) represents the vector of binary variables, and \( Q \) denotes the matrix of constants, commonly referred to as the QUBO matrix.

\begin{equation}
    minimize/maximize \quad y = x^{t}Qx \label{eq:quboform}
\end{equation}


Given that the original objective function already meets the requirements of the QUBO model, our focus at this stage is to transform the constraints into quadratic penalty terms and incorporate them into the original objective function. For each vertex \( i \in V \), the general form of Constraint \ref{const:i1} can be expressed as follows:

\begin{equation}
  X_{1} + X_{2} + \dots + X_{n} \geq 1, \quad n = |N[i]| \label{eq:c1normal}  
  \end{equation}
According to \cite{RN416}, for \( n = 1 \), the constraint can be expressed as \( P \cdot (X_{1} - 1)^{2} \), where \( P \) represents the penalty coefficient. For \( n = 2 \), the expression takes the form \( P \cdot (1 - X_{1} - X_{2} + X_{1} \cdot X_{2}) \). When \( n \geq 3 \), we must convert the inequality constraint into an equality constraint by introducing slack variables, as demonstrated in Eq. \ref{eq:sceq}.
\begin{equation}
  X_{1} + X_{2} + \dots + X_{n} - S - 1 = 0 \label{eq:sceq}
\end{equation}
It is evident that \( S \) lies within the range \( [0, n-1] \). Next, we express \( S \) as a combination of binary (0-1) variables. Given that \( S \) can assume all integer values within the interval \( [0, n-1] \), we can represent \( S \) as follows:
\begin{equation}
  S = \sum\limits_{i=1}^{bl_{n-1}-1} X_{i}^{\prime} \cdot 2^{i-1} + (n - 1 - \sum\limits_{i=1}^{bl_{n-1}-1}2^{i-1}) \cdot X_{bl_{n-1}}^{\prime} \label{eq:sc}
\end{equation}
The variable \( X_{*}^{\prime} \in \{0,1\} \) is an additional binary variable introduced to represent slack variables, while \( bl_{n} \) denotes the binary length of \( n \). We can substitute the representation from Eq. \ref{eq:sc} into Eq. \ref{eq:sceq} and square the entire expression to derive the quadratic penalty. For Constraint \ref{const:i3}, we can express it as \( P \cdot (X_{i} \cdot X_{j}) \). Assuming that for all vertices \( i \in V \), the condition \( |N[i]| \geq 3 \) holds, the QUBO model for the IDP is articulated in Eq. \ref{eq:qubo}. It is important to note that for vertices where \( |N[i]| = 1 \) or \( 2 \), we will employ alternative forms of penalty terms when addressing Constraint \ref{const:i1} (\( P \cdot (X_{1} - 1)^{2} \) or \( P \cdot (1 - X_{1} - X_{2} - X_{1} \cdot X_{2}) \)). The assumption that \( |N[i]| \geq 3 \) for all vertices in Eq. \ref{eq:qubo} is made purely for the sake of simplifying the formulation of the objective function. Consequently, Eq. \ref{eq:qubo} is provided as an illustrative example to aid the reader's understanding.
\begin{equation}
  \begin{split} 
  &\min_{\{X,X^{\prime}\}} \quad  \sum\limits_{i=1}^{|V|} X_{i}  \\
  &+ \sum\limits_{i \in V} P \cdot [\sum\limits_{j \in N[i]}X_{j}  - (\sum\limits_{i=1}^{bl_{|N[i]|-1}-1} X_{i}^{\prime} \cdot 2^{i-1} + (|N[i]| - 1 - \sum\limits_{i=1}^{bl_{|N[i]|-1}-1}2^{i-1}) \cdot X_{bl_{|N[i]|-1}}^{\prime}) - 1]^{2} \\
  &+ \sum\limits_{ij \in E}P \cdot X_{i} \cdot X_{j} \label{eq:qubo}
  \end{split}
\end{equation}

\subsection{QUBO model $\to$ Hamiltonian}
To convert the objective function in Eq. \ref{eq:qubo} into a Hamiltonian, we need to substitute the binary variables \( X_{i} \) with new variables \( s_{i} \in \{-1, 1\} \). The substitution process is outlined as follows:
\begin{equation}
  X_{i} = \frac{s_{i} + 1}{2}
\end{equation}
Then we have
\begin{equation}
  \begin{split} 
  &\min_{\{s,s^{\prime}\}} \quad  \sum\limits_{i=1}^{|V|} \frac{s_{i} + 1}{2}   \\
  &+ \sum\limits_{i \in V} P \cdot [\sum\limits_{j \in N[i]}\frac{s_{j} + 1}{2}\\  
  &- (\sum\limits_{i=1}^{bl_{|N[i]|-1}-1} \frac{s_{i}^{\prime} + 1}{2}\cdot2^{i-1} + (|N[i]| - 1 - \sum\limits_{i=1}^{bl_{|N[i]|-1}-1}2^{i-1}) \cdot \frac{s_{bl_{|N[i]|-1}}^{\prime} + 1}{2}) - 1]^{2} \\
  &+ \sum\limits_{ij \in E}P \cdot \frac{s_{i} + 1}{2} \cdot \frac{s_{j} + 1}{2} \label{eq:squbo}
  \end{split}
\end{equation}
By replacing $s$ and $s^{\prime}$ with the Pauli-Z operator $\sigma^{z}$, we ultimately obtain the Hamiltonian.
\begin{equation}
  \begin{split} 
  &H_{c} = \sum\limits_{i=1}^{|V|} \frac{\sigma_{i}^{z} + 1}{2}   \\
  &+ \sum\limits_{i \in V} P \cdot [\sum\limits_{j \in N[i]}\frac{\sigma_{j}^{z} + 1}{2}\\  
  &- (\sum\limits_{i=1}^{bl_{|N[i]|-1}-1} \frac{(\sigma_{i}^{z})^{\prime} + 1}{2}\cdot2^{i-1} + (|N[i]| - 1 - \sum\limits_{i=1}^{bl_{|N[i]|-1}-1}2^{i-1}) \cdot \frac{(\sigma_{bl_{|N[i]|-1}}^{z})^{\prime} + 1}{2}) - 1]^{2} \\
  &+ \sum\limits_{ij \in E}P \cdot \frac{\sigma_{i}^{z} + 1}{2} \cdot \frac{\sigma_{j}^{z} + 1}{2} \label{eq:h}
  \end{split}
\end{equation}

\section{QAOA}\label{sec:QAOA}
\begin{figure}[H]
  \centering
  \includegraphics[width=12cm]{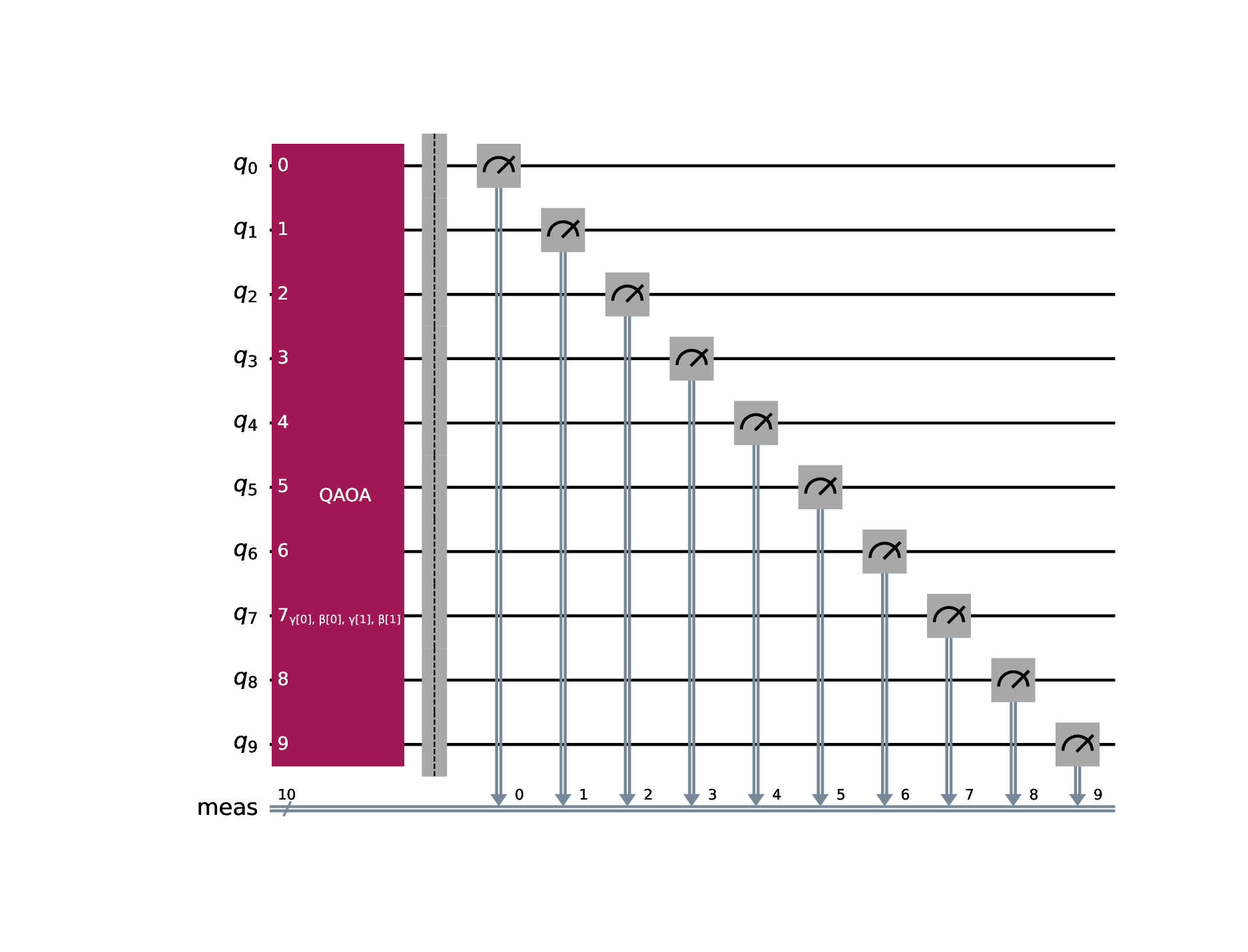}
  \caption {Basic working flow of QAOA with 2 layers of 10 qubits.}
  \label{fig:basicflow}
\end{figure}

In this section, we begin by introducing the fundamental concepts of the QAOA. For a quantum system comprising \( n \) qubits, we denote \( \ket{z} \) as a state vector in the \( 2^{n} \)-dimensional Hilbert space, where the bit string \( z \) is expressed as \( z = z_{1} z_{2} z_{3} \dots z_{n} \). The QAOA starts by applying the Hadamard gate to prepare the initial state \( \ket{s} \) from the state \( \ket{\underbrace{00 \ldots 0}_{n}} \).
\begin{equation}
  \ket{s} = \underbrace{\hat{H} \otimes \hat{H} \cdots \otimes \hat{H}}_{n} \ket{\underbrace{00\dots 0}_{n}} = \frac{1}{\sqrt{2^{n}} } \cdot \sum\limits_{z} \ket{z}
\end{equation}
Then we use two parametric unitary transformations, $U(C,\gamma)$ and $U(B,\beta)$ 
\begin{align*}
  U(C,\gamma) &= e^{-i\gamma H_{c}}\\
  U(B,\beta) &=  e^{-i\beta B}
\end{align*}
where $C = H_c$, $B = \sum\limits_{j=1}^{n} \sigma_{j}^{x}$, $\gamma \in [0,2\pi]$ and $\beta \in [0,\pi]$. By applying these two types of unitary transformations to the initial state \( \ket{s} \) and repeating the process \( q \) times, we arrive at the final state \( \ket{\gamma, \beta} \) as described in Eq. \ref{eq:gammabeta}. In this paper, \( q \) is also referred to as the layer number of the QAOA.
\begin{equation}
  \ket{\gamma,\beta} = U(B,\beta_{q})U(C,\gamma_{q}) \cdots U(B,\beta_{1})U(C,\gamma_{1}) \ket{s} \label{eq:gammabeta}
\end{equation}
After obtaining $\ket{\gamma,\beta}$, we can measure the expectation value of $H_{c}$.
\begin{equation}
  F_{q}(\gamma,\beta) = \bra{\gamma,\beta} H_{c} \ket{\gamma,\beta}
\end{equation}
As a reference, Fig. \ref{fig:basicflow} shows the quantum circuit for 10 qubits with 2 layers.
For the IDP in this paper, our task is to find such $\gamma$ and $\beta$ with the minimal expectation value of $F_{q}(\gamma, \beta)$. The QAOA is a hybrid algorithm that combines a classical optimizer with a quantum computer. We can use a classical optimizer to search for such $\gamma$ and $\beta$. Next, we summarize the steps of using the QAOA to solve the IDP.
\begin{enumerate}[label=Step \arabic*:]
  \item Convert the IDP into a QUBO model.
  \item Transform the QUBO model into a Hamiltonian.
  \item Set the layer number $q$ and convert the Hamiltonian into a quantum circuit. IBM's Qiskit integrates many tools, such as QAOAAnsatz. We can generate the quantum circuit by inputting the Hamiltonian and the layer number.
  \item Initialize $\gamma$ and $\beta$ for each layer.
  \item Use a classical optimizer, such as COBYLA, to optimize $\gamma$ and $\beta$. The optimization process terminates when the maximum number of iterations is reached or the predefined function tolerance is satisfied, yielding $\gamma_{*}$ and $\beta_{*}$.
  \item Update the quantum circuit with $\gamma_{*}$ and $\beta_{*}$.
  \item Perform multiple samplings on the updated quantum circuit and output the bit string $z$ with the highest probability.
  \item Derive the IDS of the graph based on $z$.
\end{enumerate}

\section{Experiment}\label{sec:Experiment}
\begin{figure}[H]
  \centering
  \includegraphics[width=12cm]{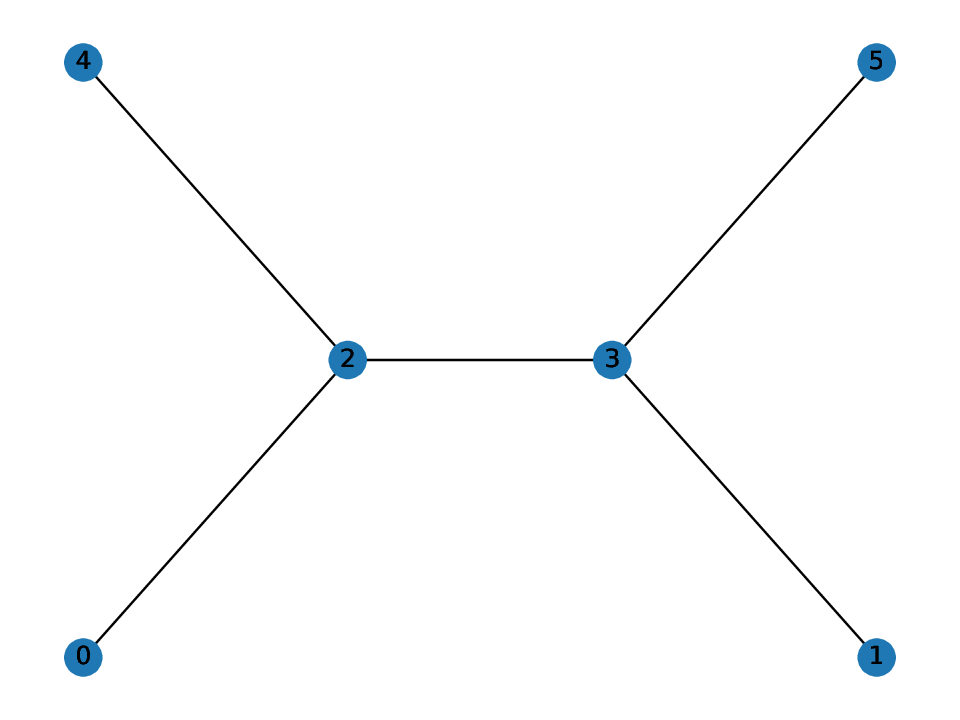}
  \caption {A 6 nodes graph.}
  \label{fig:6nodegraph}
\end{figure}

 In this chapter, we will evaluate the performance of the QAOA-based algorithm in solving the IDP. The graph selected for this analysis is depicted in Fig. \ref{fig:6nodegraph}. This graph comprises 6 vertices, and it is noteworthy that the optimal dominating set and the optimal IDS differ. Specifically, the optimal dominating set is \( \{2, 3\} \), while the optimal IDS, due to the symmetry of the graph, presents two possible solutions: \( \{0, 4, 3\} \) or \( \{1, 5, 2\} \).
Based on the modeling method outlined in Section \ref{sec:problemmodeling}, the QUBO model of the IDP for this graph is presented in Eq. \ref{eq:expqubo}. The corresponding Hamiltonian for this model can be readily obtained by replacing \( x_{*} \) with \( \frac{s_{*} + 1}{2} \) and substituting \( s_{*} \) with \( \sigma_{*}^{z} \). For the sake of brevity, the detailed expansion of this process is omitted here.
\begin{equation}
  \begin{split}
    minimize \quad & x_{0} + x_{1} + x_{2} + x_{3} + x_{4} + x_{5} \\
& + P \cdot \left(1 - x_{0} - x_{2} + x_{0} x_{2}\right) \\
& + P \cdot \left(1 - x_{1} - x_{3} + x_{1} x_{3}\right) \\
& + P \cdot \left(x_{2} + x_{0} + x_{4} + x_{3} - \left(x_{6} + 2x_{7}\right) - 1\right)^{2} \\
& + P\cdot\left(x_{3} + x_{1} + x_{2} + x_{5} - \left(x_{8} + 2x_{9}\right) - 1\right)^{2} \\
& + P\cdot \left(1 - x_{4} - x_{2} + x_{4} x_{2}\right) \\
& + P \cdot\left(1 - x_{5} - x_{3} + x_{5} x_{3}\right) \\
& + P \cdot x_{0} x_{2} + P \cdot x_{1} x_{3} + P \cdot x_{2} x_{4} + P \cdot x_{2} x_{3} + P \cdot x_{3} x_{5} \label{eq:expqubo}
  \end{split}
\end{equation}
In this experiment, we utilized IBM's qasm_simulator as the backend, with QAOAAnsatz serving as the quantum circuit generator and BaseSamplerV1 as the sampler. The optimization method employed was COBYLA, configured with a default function tolerance of \( 10^{-8} \). The cost function utilized for the optimization was the Conditional Value-at-Risk (CVAR) \citep{RN457}. For the initial values of \( \gamma \) and \( \beta \) in the QAOA, we adopted the initialization method proposed by \cite{RN458}. The key parameters considered in the experiment include the layer number \( q \) of the QAOA, the \( alpha \) parameter of the CVAR, the penalty coefficient \( P \) of the QUBO model, and the maximum iterations allowed for COBYLA.

The experiment is divided into two main parts: fundamental testing and robustness testing. In the fundamental testing phase, we evaluate the performance of the QAOA in solving the IDP using specific parameters. The robustness testing phase involves comparing the computational results of the QAOA under varying parameter settings. Additionally, this chapter includes an analysis of the complexity of the QAOA.

\subsection{fundamental testing} \label{sec:fundamental}

\begin{figure}[H]
  \centering
  \includegraphics[width=12cm]{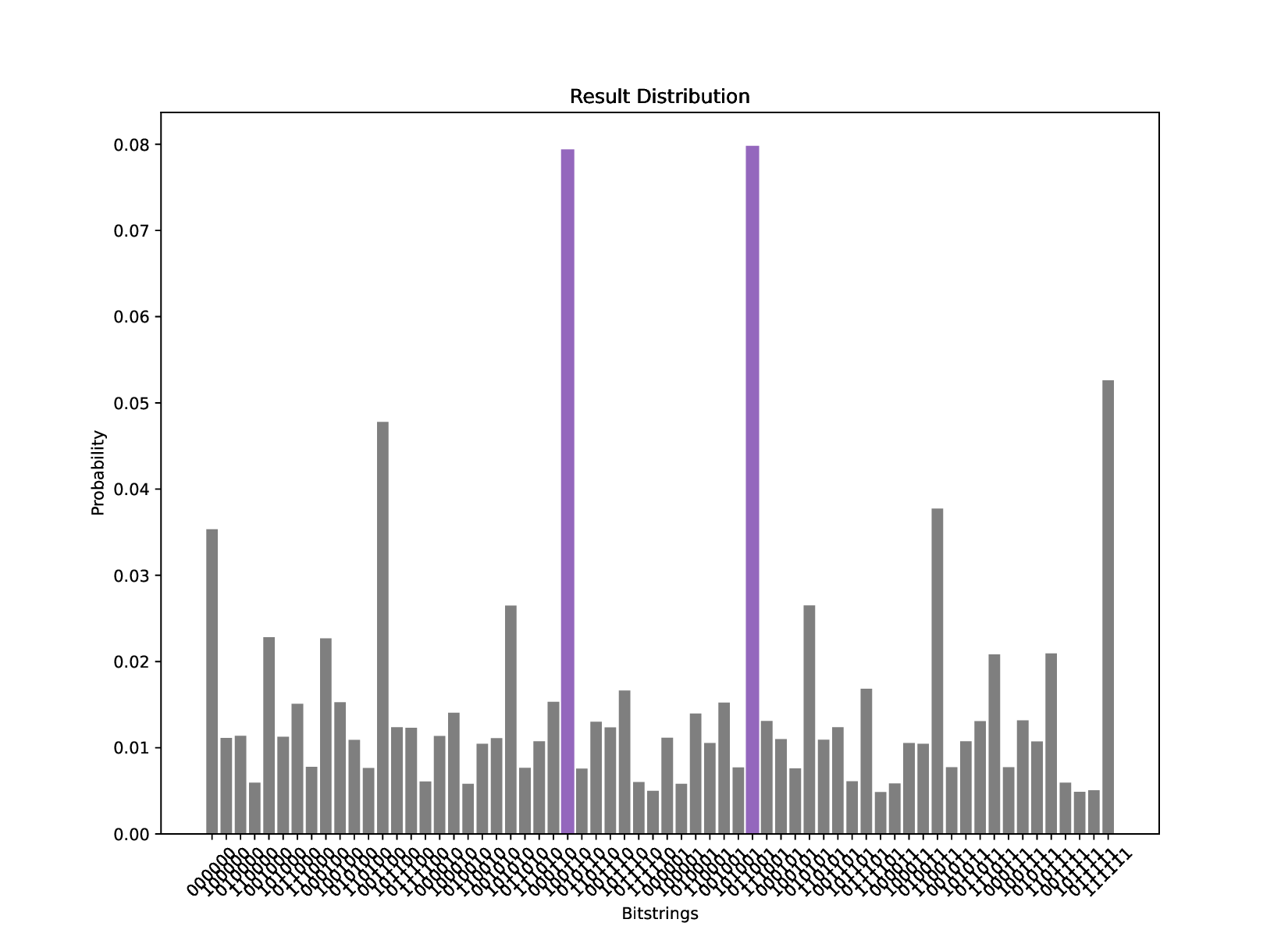}
  \caption {Distribution of the bit strings.}
  \label{fig:distribution}
\end{figure}

\begin{figure}[H]
  \centering
  \includegraphics[width=11cm]{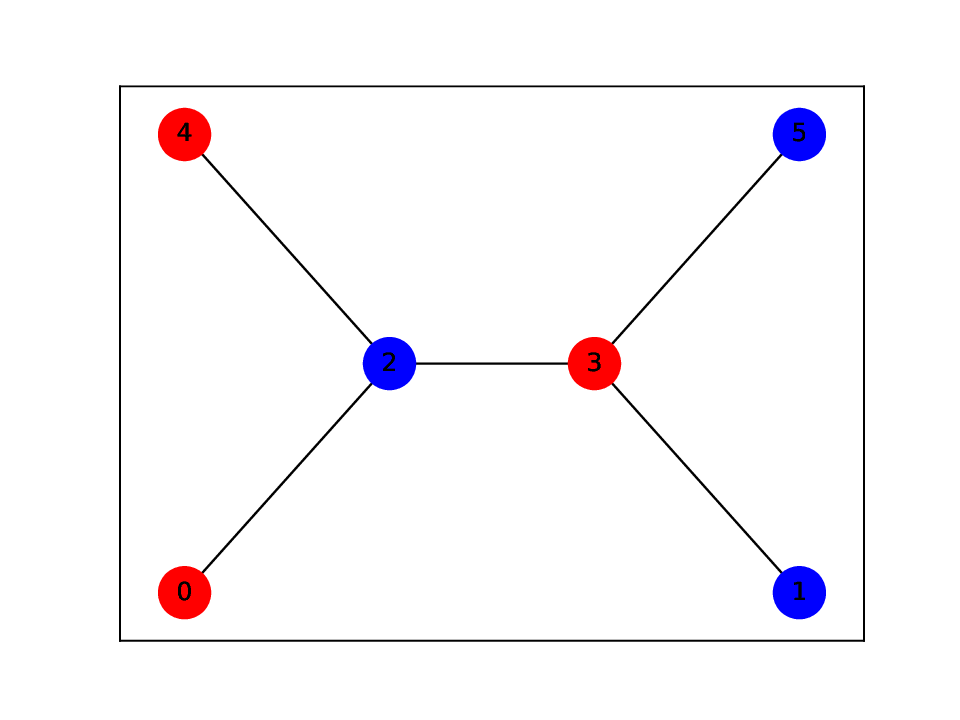}
  \caption {Graph with IDS $\{1,2,5\}$. }
  \label{fig:d1opt}
\end{figure}

\begin{figure}[H]
  \centering
  \includegraphics[width=11cm]{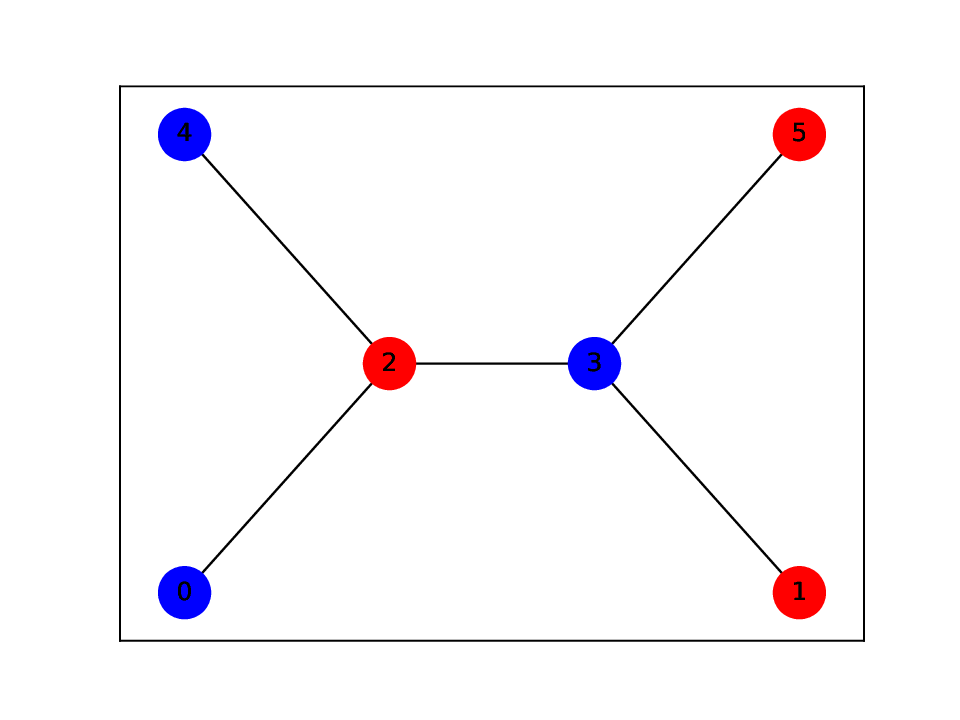}
  \caption {Graph with IDS $\{0,3,4\}$. }
  \label{fig:d2opt}
\end{figure}

We set the parameters to \( q = 15 \), \( alpha = 0.3 \), and \( P = 4.5 \), which corresponds to 0.75 times the number of nodes in the selected graph, with a maximum of 10,000 iterations. Under these conditions, we employed the QAOA to solve the IDP for the 6-node graph illustrated in Fig. \ref{fig:6nodegraph}. Figure \ref{fig:distribution} presents the distribution of the final sampling results, indicating that the probabilities of the bit strings \( z = 011001 \) and \( z = 100110 \), which are 0.0797 and 0.0793 respectively, are significantly higher than those of other configurations. Visualizing the results for \( 011001 \) and \( 100110 \) on the graph reveals that this outcome arises from the inherent symmetry of the graph (as shown in Figs. \ref{fig:d1opt} and \ref{fig:d2opt}). This finding clearly demonstrates the effectiveness of the QAOA algorithm in solving the IDP.

\begin{figure}[H]
  \centering
  \includegraphics[width=12cm]{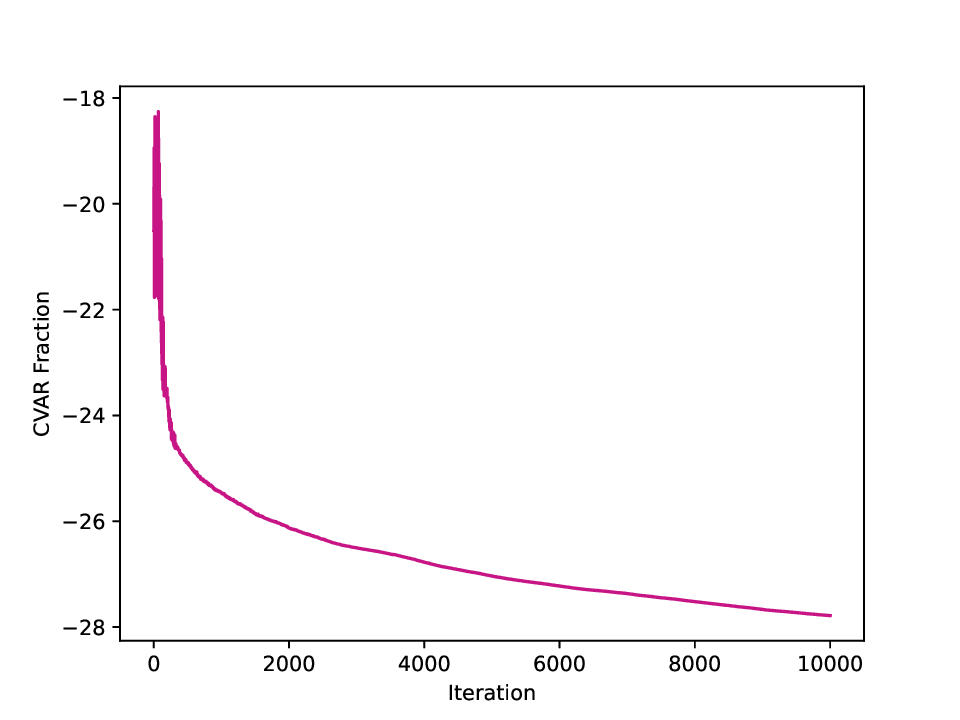}
  \caption {Cost of QAOA.}
  \label{fig:cost}
\end{figure}

In Fig. \ref{fig:cost}, we record the cost as it evolves over the course of the iterations. We observe a sharp decrease in cost within the range of \( [0, 500] \) iterations, followed by a more gradual decline thereafter. One possible explanation for this behavior is that the value of the penalty term surpasses that of the original objective function. The significant reduction in the cost during the \( [0, 500] \) interval may be attributed to the optimizer initially searching for bit strings \( z \) that render the penalty terms in the objective function equal to zero. For instance, in extreme cases such as the largest IDS of the graph, \( \{0, 1, 4, 5\} \), the corresponding bit string is \( z = 110011 \). Once the penalty terms are minimized, the optimization process shifts its focus to finding bit strings that minimize the size of the IDS. It is important to note that, although the cost decrease becomes more gradual after 500 iterations, it continues to exhibit a clear downward trend. However, this gradual decline does not impact the fact that \( z = 011001 \) and \( z = 100110 \) remain the two bit strings with the highest probabilities. Further exploration may enhance the probabilities of obtaining correct and optimal results; at the current parameter settings, these probabilities are 0.197 and 0.159, respectively, as indicated in Table \ref{table:defdsps}. Here, the correct probability refers to the likelihood that the sampled result from the QAOA corresponds to an IDS, while the optimal probability denotes the likelihood of obtaining the optimal IDS.

\begin{longtable}{|c|c|}
  \caption{The correct and optimal probability}
  \label{table:defdsps}\\
  \hline
  Correct & Optimal \\
  \hline
  \endfirsthead
  \multicolumn{2}{r}{Continued}\\
  \hline
  Correct & Optimal \\
  \hline
  \endhead
  \hline
  \multicolumn{2}{r}{Continued on next page}\\
  \endfoot
  \endlastfoot
  \hline
  0.197 &0.159\\
  \hline
\end{longtable}

\subsection{robustness testing} \label{sec:robustness}

\begin{figure}[H]
  \centering
  \includegraphics[width=12cm]{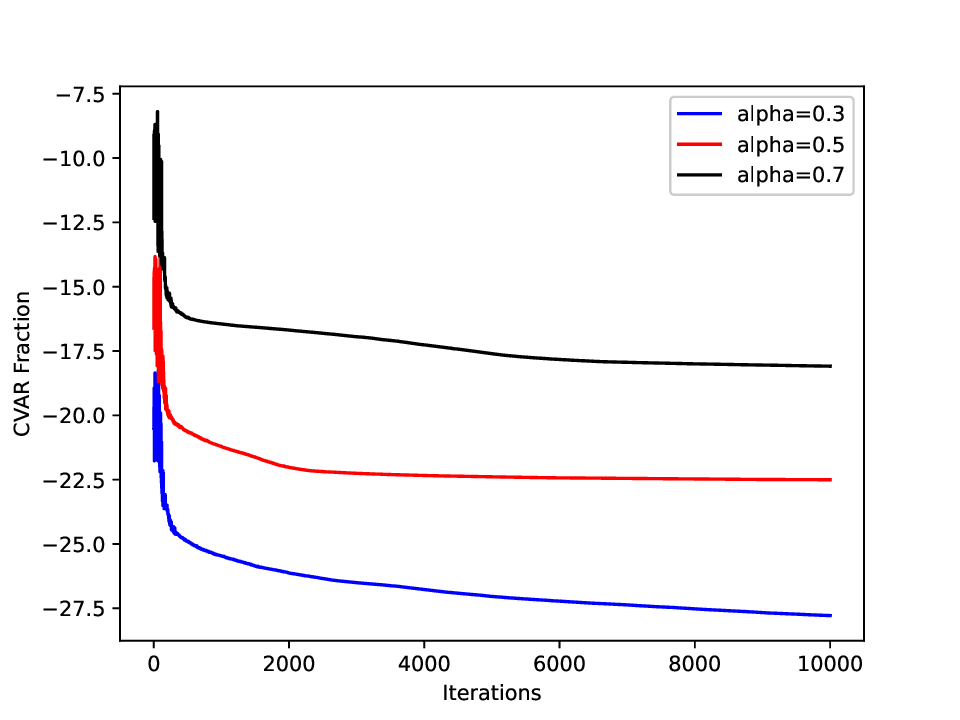}
  \caption {Cost of QAOA with different $alpha$s.}
  \label{fig:costcompare}
\end{figure}

In this section, we conduct a robustness analysis on the four parameters of the QAOA algorithm: \( alpha \), \( q \), \( P \), and the maximum number of iterations. Initially, in Fig. \ref{fig:costcompare}, we set \( q = 15 \), \( P = 4.5 \), and a maximum of 10000 iterations, and we compare the changes in cost for \( alpha = 0.3 \), \( 0.5 \), and \( 0.7 \). It is evident that, consistent with our observations in Fig. \ref{fig:cost}, the cost decreases sharply within the range of \( [0, 500] \) iterations for all values of \( \alpha \). Beyond this range, the reduction in cost becomes more gradual. Overall, the trend of cost reduction remains consistent across the different levels of \( alpha \).

In Tables \ref{table:difflayer}, \ref{table:diffpunish}, and \ref{table:diffmaxiter}, we adjust the layer number, penalty coefficient, and maximum iterations, respectively, to observe their effects on the correct and optimal probabilities.

In Table \ref{table:difflayer}, we observe that as the number of layers increases, the correct probability exhibits an upward trend. However, the optimal probability does not consistently increase with \( q \), contrary to theoretical expectations \citep{RN436}. For instance, the optimal probability at \( q = 20 \) is 0.135, while at \( q = 15 \), it is 0.159. We believe this discrepancy arises because the COBYLA optimizer requires more iterations when \( q = 20 \), given that the number of \( \gamma \) and \( \beta \) parameters increases by 10 compared to \( q = 15 \). This situation underscores the complexity of the IDP. Nevertheless, it is noteworthy that the optimal probabilities for \( q = 15 \) and \( q = 20 \) are significantly higher than those for \( q = 10 \).

In Table \ref{table:diffpunish}, we compare the impact of different penalty coefficients on the correct and optimal probabilities. The four values of \( P \) correspond to 0.5, 0.75, 1, and 1.5 times the number of vertices, which serves as an upper bound for the size of the IDS. Consequently, these parameters reflect varying proportions between the penalty term and the original objective function. Our analysis reveals no clear correlation between the value of \( P \) and the two probability metrics. Moreover, even for two closely related penalty coefficients, such as \( P = 3 \) and \( P = 4.5 \), we observe significant differences in their corresponding correct and optimal probabilities. This finding underscores the necessity for careful selection of \( P \), as also emphasized in \cite{RN416}. The recommendation from \cite{RN416} is to initially set \( P \) between 0.75 and 1.5 times the original objective function, which aligns well with our experimental results.

Next, based on the data presented in Table \ref{table:diffmaxiter}, we observe that as the number of maximum iterations increases, both the correct and optimal probabilities show a clear upward trend, which aligns with our expectations.

In addition to these analyses, we tested a total of 144 parameter combinations, varying \( q \in \{10, 15, 20\} \), \( alpha \in \{0.3, 0.5, 0.7\} \), \( P \in \{3, 4.5, 6, 9\} \), and the maximum iterations \( \in \{100, 500, 1000, 10000\} \). Across these parameter settings, the two most probable bit strings, \( 011001 \) and \( 100110 \), appeared 21 times. Furthermore, one of these two bit strings ranked among the top two in 54 instances. This finding suggests that the QAOA algorithm is effective in solving the IDP under specific parameter configurations.

Based on these results, we can conclude that the solution quality of the algorithm is significantly influenced by the selected parameters, and optimal performance is not guaranteed across every combination. Our experiences of extensive numerical experiments indicate that solving the IDP usually requires more than 10 layers, with the most effective solutions typically arising when the number of maximum iterations exceeds 500.

\begin{longtable}{|c|c|c|}
  \caption{The comparison of correct and optimal probabilities of different layers when $alpha = 0.3$, $P = 4.5$ and maximal iterations = 10000.}
  \label{table:difflayer}\\
  \hline
  \diagbox{Layer number}{Probability}& Correct & Optimal \\
  \hline
  \endfirsthead
  \multicolumn{3}{r}{Continued}\\
  \hline
  \diagbox{Layer number}{Probability}& Correct & Optimal \\
  \hline
  \endhead
  \hline
  \multicolumn{3}{r}{Continued on next page}\\
  \endfoot
  \endlastfoot
  \hline
  $q = 10$&0.177 & 0.077\\
  \hline
  $q = 15$&0.197 & 0.159\\
  \hline
  $q = 20$&0.201 & 0.135\\
  \hline
\end{longtable}

\begin{longtable}{|c|c|c|}
  \caption{The comparison of correct and optimal probabilities of different punishment coefficients when $q=15$, $alpha = 0.3$ and maximal iterations = 10000.}
  \label{table:diffpunish}\\
  \hline
  \diagbox{Punishment coefficient}{Probability}& Correct & Optimal \\
  \hline
  \endfirsthead
  \multicolumn{3}{r}{Continued}\\
  \hline
  \diagbox{Punishment coefficient}{Probability}& Correct & Optimal \\
  \hline
  \endhead
  \hline
  \multicolumn{3}{r}{Continued on next page}\\
  \endfoot
  \endlastfoot
  \hline
  $P = 3$&0.098 & 0.067\\
  \hline
  $P = 4.5$&0.197 & 0.159\\
  \hline
  $P = 6$&0.196 & 0.135\\
  \hline
  $P = 9$&0.176 & 0.102\\
  \hline
\end{longtable}

\begin{longtable}{|c|c|c|}
  \caption{The comparison of correct and optimal probabilities of different maximal iterations when $q=15$, $alpha = 0.3$ and $P = 4.5$.}
  \label{table:diffmaxiter}\\
  \hline
  \diagbox{Maximal  iterations}{Probability}& Correct & Optimal \\
  \hline
  \endfirsthead
  \multicolumn{3}{r}{Continued}\\
  \hline
  \diagbox{Maximal  iterations}{Probability}& Correct & Optimal \\
  \hline
  \endhead
  \hline
  \multicolumn{3}{r}{Continued on next page}\\
  \endfoot
  \endlastfoot
  \hline
  $100$&0.100 & 0.057\\
  \hline
  $500$&0.144 & 0.088\\
  \hline
  $1000$&0.149 & 0.094\\
  \hline
  $10000$&0.197 & 0.159\\
  \hline
\end{longtable}

\subsection{complexity analysis} \label{sec:complex}

\begin{longtable}{|c|c|c|}
  \caption{The comparison of time complexity of algorithms for IDP.}
  \label{table:diffcom}\\
  \hline
   Algorithm & Time complexity & Support quantum computer? \\
  \hline
  \endfirsthead
  \multicolumn{3}{r}{Continued}\\
  \hline
  Algorithm & Time complexity & Support quantum computer? \\
  \hline
  \endhead
  \hline
  \multicolumn{3}{r}{Continued on next page}\\
  \endfoot
  \endlastfoot
  \hline
  QAOA & $O[poly(q) + poly(m)]$ & Yes\\
  \hline
  \cite{RN442}& $O[|V|^{2}]$ & No\\
  \hline
  \cite{RN443}& $O[|V|^{2.376}]$ & No\\
  \hline
  \cite{RN441}(for (k,l)-graphs)& $O[3.3028^{k} + |V|]$ & No\\
  \hline
  \cite{RN451} & $O[2^{0.529|V|}]$ & No\\
  \hline
  \cite{RN452} & $O[2^{0.441|V|}]$ & No\\
  \hline
  \cite{RN450} & $O[2^{0.424|V|}]$ & No\\
  \hline
\end{longtable}
 
\cite{RN334} analyzed the time complexity of the QAOA. The time complexity consists of two components. The first component, \( O[\text{poly}(q)] \), represents the most time-consuming aspect of the QAOA, which stems from its multi-layer structure. The second component pertains to the time complexity of the optimization algorithm utilized. In this paper, we employ the COBYLA as the optimization algorithm, which has a time complexity of \( O[m] \), where \( m \) represents the maximum number of iterations. Therefore, the overall time complexity of the QAOA can be expressed as \( O[\text{poly}(q) + \text{poly}(m)] \).

In Table \ref{table:diffcom}, we compare the time complexity of the QAOA with other algorithms previously employed for solving the IDP. Our analysis reveals that QAOA provides a significant advantage in terms of time complexity when addressing the IDP, and can be executed on a quantum computer. However, this does not imply that QAOA is superior in all aspects for solving the IDP. Based on the experimental results and analyses presented in Sections \ref{sec:fundamental} and \ref{sec:robustness}, we conclude that QAOA is effective for solving the IDP, provided that the algorithm's parameters are carefully selected. Moreover, there is still room for improvement regarding both the accuracy of the results and the probability of obtaining the optimal solution.

Additionally, when executing the QAOA algorithm on classical computers, it is crucial to consider memory constraints. This requirement arises because, during the sampling process for a combinatorial optimization problem involving \( n \) qubits, we must locally store a matrix with a length of \( 2^{n} \) and a width corresponding to the number of samples per bit string. Consequently, substantial memory overhead is necessary when running the QAOA on a local quantum simulator.

\section{Conclusion}\label{sec:conclusion}

In this paper, we explored the use of the quantum algorithm, QAOA, to solve the IDP. We first modeled the IDP as a 0-1 integer programming problem and transformed different constraints into corresponding quadratic penalties, which were added to the original objective function, resulting in the QUBO model of the IDP. The QUBO model was then converted into a Hamiltonian, completing the preparation for inputting the IDP into the quantum algorithm. We employed the QAOA algorithm to solve the IDP. This algorithm is a hybrid approach combining quantum computation with classical optimizers. By adjusting the $\gamma$ and $\beta$ parameters for each layer of QAOA through classical optimization, we gradually improved the expectation value of $H_{c}$, which corresponds to the objective function of the IDP. 

In the experimental section, we conducted both fundamental and robustness testing. We found that the QAOA algorithm is effective for solving the IDP, as the final sampling results reflected multiple IDSs due to the inherent symmetry of the graph. From the complexity analysis, we confirmed that the QAOA algorithm offers advantages in terms of complexity compared to other algorithms for solving the IDP. However, based on the robustness testing, we concluded that the effectiveness of the algorithm depends on the selected parameters. The QAOA algorithm cannot guarantee correct solutions for the problem under arbitrary parameter combinations, which is one of the limitations of the algorithm. Additionally, we did not optimize the quantum circuits used in this study; instead, we utilized circuits generated by IBM's Qiskit package. 

In the future, based on the work presented in this paper, our research can be extended in the following directions: 
(1) Testing the performance of the QAOA algorithm for solving the IDP on real quantum computers. (2) Explore the QAOA parameter range that is more suitable for IDP. (3) Applying the QAOA algorithm to solve other variants of the domination problem, such as the total domination problem, the perfect domination problem, and so forth.

\section*{Acknowledgement}
This work was inspired by the GitHub project named qopt-best-practices, which can be accessed at [https://github.com/qiskit-community/qopt-best-practices.git]. Additionally, we express our gratitude to IBM for providing the Qiskit package, which offers essential tools for quantum programming.

\section*{Data Availability}
The data used to support the findings of this study are included within the article.

\section*{Conflicts of interest}
The authors declare that they have no conflicts of interest that could have appeared to influence the work reported in this paper.

\section*{Funding statement}
 
This work is supported by National Natural Science Foundation of China (No. 12331014).

\section*{Declaration of Generative AI}

During the preparation of this work the authors used chatgpt in order to improve readability and language. After using this tool, the authors reviewed and edited the content as needed and take full responsibility for the content of the publication.



 \bibliographystyle{elsarticle-harv} 
 \bibliography{idom}






\end{CJK}
\end{document}